\documentclass[a4paper,11pt]{article}

\usepackage{jheppub} 

\usepackage[T1]{fontenc} 
\usepackage[dvipsnames]{xcolor}

\usepackage{booktabs}
\usepackage{float}
\usepackage{titlesec}
\usepackage{capt-of}
\usepackage{slashed,bm}
\usepackage{booktabs}

\usepackage{mathrsfs}
\usepackage{amsmath}
\usepackage{tikz}
\usepackage[com pat=1.1.0]{tikz-feynman}
\usetikzlibrary{calc}
\usepackage{slashed}
\usepackage{MnSymbol}
\usepackage{dsfont}
\usepackage{mathtools}
\usepackage{wrapfig}
\usepackage{bbold}

\newcommand{\ud}{\mathrm{d}}

\newcommand{\Lbar}{\bar{\Lambda}}
\newcommand{\gamE}{\gamma_{\text{E}}}
\newcommand{\Euler}{\mathrm{e}}
\newcommand{\as}{\alpha_s}

\newcommand{\LbarQCD}{\Lbar_{\text{{\tiny QCD}}}}
\newcommand{\LbarQED}{\Lbar_{\text{{\tiny QED}}}}
\newcommand{\muQED}{\mu_\text{\tiny QED}}

\newcommand{\LambdaIR}{\bar{\Lambda}_{\text{\tiny QED}}}

\newcommand{\muB}{\mu_\text{B}}
\newcommand{\muq}{\mu_\text{q}}

\DeclareMathOperator*{\SumInt}{%
\mathchoice%
  {\ooalign{$\displaystyle\sum$\cr\hidewidth$\displaystyle\int$\hidewidth\cr}}
  {\ooalign{\raisebox{.14\height}{\scalebox{.7}{$\textstyle\sum$}}\cr\hidewidth$\textstyle\int$\hidewidth\cr}}
  {\ooalign{\raisebox{.2\height}{\scalebox{.6}{$\scriptstyle\sum$}}\cr$\scriptstyle\int$\cr}}
  {\ooalign{\raisebox{.2\height}{\scalebox{.6}{$\scriptstyle\sum$}}\cr$\scriptstyle\int$\cr}}
}

\title{Static linear response of hot and dense QCD matter to electromagnetic fields: Leading hard and soft QCD corrections}

\renewcommand*{\thefootnote}{\fnsymbol{footnote}}
\author[a, b]{Osvaldo Ferreira,}
\author[b]{Eduardo S. Fraga,}
\author[c,d]{Tyler Gorda,}
\author[e,f]{Risto Paatelainen,}
\author[f]{Leon Sandbote}
\author[g]{and Kaapo Seppänen}

\affiliation[a]{
 Instituto de F\'isica, Universidade de S\~ao Paulo, R. do Mat\~ao, 1371, S\~ao Paulo, Brazil, 05508-090
}
\affiliation[b]{Instituto de F\'\i sica, Universidade Federal do Rio de Janeiro,CEP 21941-972 Rio de Janeiro, RJ, Brazil}

\affiliation[c]{Center for Cosmology and AstroParticle Physics (CCAPP), Ohio State University, Columbus, OH 43210, USA}
\affiliation[d]{Department of Physics, The Ohio State University, Columbus, OH 43210, USA}
\affiliation[e]{Department of Physics and Astronomy, FI-20014 University of Turku, Finland}
\affiliation[f]{Department of Physics and Helsinki Institute of Physics, P.O.~Box 64, FI-00014 University of Helsinki, Finland}
\affiliation[g]{AEC, Institute for Theoretical Physics, University of Bern, Sidlerstrasse 5, CH-3012 Bern, Switzerland}

\emailAdd{osvaldof@if.usp.br}
\emailAdd{fraga@if.ufrj.br}
\emailAdd{gorda.1@osu.edu}
\emailAdd{risto.paatelainen@utu.fi}
\emailAdd{leon.sandbote@helsinki.fi}
\emailAdd{kaapo.seppaenen@unibe.ch}

\abstract{
We compute the static electromagnetic susceptibilities of a hot and dense quark-gluon plasma using perturbative Quantum Chromodynamics (QCD). Our evaluation  includes the leading $\mathcal{O}(\as)$ correction as well as the leading soft, resummed contribution of $\mathcal{O}(\as^{3/2})$ within electrostatic QCD. By matching to Lattice QCD at vanishing baryon chemical potential through Lattice perturbation theory, we establish a connection between perturbative results and Lattice simulations and assess the size of higher-order corrections. This extends the electromagnetic susceptibilities to finite baryon chemical potential, where Lattice methods are not applicable and establishes first-principle constraints on the quark-gluon plasma's electromagnetic response at temperatures and densities relevant for intermediate-energy heavy-ion collisions.
}

\begin{document}
\maketitle
\flushbottom

\renewcommand*{\thefootnote}{\arabic{footnote}}

\section{Introduction}

Understanding the response of Quantum Chromodynamics (QCD) matter to background electromagnetic fields provides insight into a wide range of physical systems, from heavy-ion collisions~\cite{Kharzeev:2007jp,Skokov:2009qp,Voronyuk:2011jd,Bzdak:2011yy,Deng:2012pc,Tuchin:2013apa,Tuchin:2015oka,Inghirami:2016iru,Roy:2017yvg, Shen:2025unr} to strongly magnetized neutron stars \cite{Duncan:1992hi,Thompson:1993hn,Kouveliotou:1998ze} and their mergers~\cite{Aguilera-Miret:2025nts}, as well as the early universe~\cite{Vachaspati:1991nm,Enqvist:1993np,Grasso:2000wj}. In particular, the linear response of the medium, encoded in electromagnetic susceptibilities, 
determines how external fields influence the properties of strongly interacting matter.

The magnetic susceptibility of hot QCD matter has been extensively studied and has direct phenomenological implications in non-central heavy-ion collisions. In particular, it leads to the paramagnetic squeezing of the quark-gluon plasma, manifested as an elongation along the direction of the magnetic field~\cite{Bali:2013owa, Pang:2016yuh}, and may leave an imprint on the elliptic flow $v_2$ of charged pions~\cite{Pang:2016yuh, Jiang:2024bez}. In contrast, the electric susceptibility has received comparatively less attention, partly because the response to an external electric field involves both charge redistribution 
and genuine polarization effects that must be separated. The former leads to an infrared divergence associated with screening effects, making the isolation of the polarization contribution nontrivial~\cite{Endrodi:2022,Endrodi:2026kmb}. 
Recent Lattice studies~\cite{Endrodi:2023wwf} have succeeded in isolating this contribution and indicate that QCD matter exhibits a negative electric susceptibility.

From a theoretical perspective, electromagnetic susceptibilities have been investigated using non-perturbative Lattice methods~\cite{Bonati:2013lca,Bali:2012jv,Bali:2014kia,Bali:2020bcn} and effective models such as the Hadron Resonance Gas~\cite{Endrodi:2026kmb, Samanta:2025xjm}, linear sigma models~\cite{Tawfik:2017cdx, Samanta:2025xjm}, and the off-shell Parton-Hadron-String Dynamics model~\cite{Steinert:2013fza}. While these approaches provide valuable insight, a first-principles understanding across the QCD phase diagram remains incomplete. In particular, Lattice QCD provides reliable results at finite temperature $T$ and vanishing baryon chemical potential $\muB$, thereby offering a crucial non-perturbative benchmark. However, due to the Sign Problem \cite{Hands:2007by,deforcrand2009pos,Nagata:2021ugx}, Lattice calculations cannot be extended to large baryon density, leaving this region of the phase diagram mostly unconstrained from first-principles.

This limitation is particularly relevant for intermediate-energy heavy-ion collision programs~\cite{Galatyuk:2019lcf}, such as the Beam Energy Scan at RHIC and experiments at SPS, NICA, and FAIR, which probe regions with nonzero $\muB$. At nonzero $\muB$, perturbative QCD (pQCD) provides a systematically improvable, first-principles framework at sufficiently high temperatures and densities, where the strong coupling $\as$ is small. At vanishing baryon density, where Lattice QCD results are reliable, this enables a direct comparison that tests the convergence of the perturbative expansion and quantifies the importance of higher-order corrections.  
Once validated at vanishing density, pQCD can be used to estimate the impact of baryon density on electromagnetic observables. 

Despite their importance, electromagnetic susceptibilities have not been determined for QCD matter beyond leading order (LO)~\cite{Endrodi:2022,Ferreira:2023cqw}. Existing analytical results are restricted to LO calculations, where QCD interactions are neglected, and asymptotic high-temperature estimates. Higher-order contributions, including the leading $\mathcal{O}(\as)$ corrections, have so far only been inferred indirectly from asymptotic considerations~\cite{Endrodi:2023wwf,Bali:2020bcn,Bali:2014kia}, such as the logarithmic high-temperature behavior governed by the beta function of Quantum Electrodynamics (QED). This approach constrains the size of QCD corrections through the leading logarithmic term but does not provide a complete perturbative determination. 
While this asymptotic behavior correctly captures the leading logarithmic structure, it is insufficient for quantitative phenomenology and does not provide access to the dependence of the result on finite baryon chemical potentials.

In this work, we compute the electromagnetic susceptibilities of hot and dense QCD matter in perturbation theory, including the leading QCD correction at $\mathcal{O}(\as)$ and the leading soft, resummed contribution at $\mathcal{O}(\as^{3/2})$ within the dimensionally-reduced effective theory of Electrostatic QCD (EQCD)~\cite{GINSPARG1980388,PhysRevD.27.917,PhysRevD.23.2305}, which provides a sizable correction. We first benchmark our perturbative results against Lattice QCD at vanishing $\muB$, establishing a consistent connection between continuum perturbation theory and Lattice renormalization schemes. This is achieved through a calculation using Lattice perturbation theory (LPT), which isolates the vacuum contribution and determines the corresponding vacuum subtraction scale. Having validated the approach at zero density, we then extend the calculation to finite $\muB$, where Lattice methods are not applicable. Our results therefore provide the first controlled first-principles determination of electromagnetic susceptibilities at finite baryon chemical potential, bridging perturbative and non-perturbative approaches and offering quantitative insight relevant for current and future heavy-ion collision experiments, besides possible astrophysical applications in magnetized neutron stars and their mergers. 

This paper is organized as follows. In Sec.~\ref{sec:static-response} we define and discuss the electromagnetic susceptibilities and their renormalization. In Sec.~\ref{sec:summary-calc} we present a summary of our calculation, including the LO hard contribution, the LO soft contribution, and the vacuum-contribution matching. In Sec.~\ref{sec:summary-results} we present our main results, with direct comparison to Lattice data in the case of vanishing baryon chemical potential and predictions for nonzero $\muB$. Sec.~\ref{conclusions} contains our conclusions, and a few relevant technical results can be found in the Appendices.

\section{Static response of a medium to background electromagnetic fields}
\label{sec:static-response}

\subsection{Electromagnetic susceptibilities}

The electromagnetic susceptibilities quantify the linear response of a medium to an external background electromagnetic field. In this work, we are interested in the static (i.e. $k_0=0$) and homogeneous (i.e. $k \rightarrow 0$) limit, with $K\equiv \left(k_0, k \right)$ the external Euclidean photon four-momentum. The bare susceptibilities can be obtained by expanding the photon polarization tensor 
\begin{equation}
\Pi_{\mu \nu}(K)  \ = \
\begin{tikzpicture}[baseline=0.0cm]
    \begin{feynman}
    \vertex at (0,0) (f0);
    \vertex at (1.2,0) (f2);
    \vertex at (-0.6,0) (p0);
    \vertex at (1.8,0) (p1);
    \vertex[blob] (b)at (0.6,0) {};
    \diagram*{
    (p0) --[photon, edge label' = \(\mu\), momentum =\(K\)] (b) --[photon, edge label' = \(\nu\),  momentum =\(K\)] (p1);
    };
    \end{feynman}
\end{tikzpicture}
\end{equation}
with respect to $k$~\cite{Weldon:1982}. 
In fact, 
the magnetic susceptibility $\chi_b$ is related to the spatial part of the photon polarization tensor~\cite{Weldon:1982},
\begin{equation}\label{eq: bare magnetic susceptibility}
        \chi_b \, =\, \frac{1}{2(d-1)e^2} 
        \lim _{k \rightarrow 0} \frac{\partial^2 \Pi_{ii}}{\partial k^2}\left(0, k\right),
\end{equation}
while the bare electric susceptibility $\xi_b$ is given in terms of the temporal component~\cite{Endrodi:2022}
\begin{equation}\label{eq: bare electric susceptibility}
     \xi_b \,=\, -\frac{1}{2e^2}
     \lim _{k \rightarrow 0} \frac{\partial^2 \Pi_{00}}{\partial k^2} \left(0, k\right) .
\end{equation}

As discussed in Refs.~\cite{Endrodi:2022, Ferreira:2023cqw, Endrodi:2026kmb}, the electric susceptibility evaluated using Eq.~\eqref{eq: bare electric susceptibility} quantifies the polarization of the medium disentangled from the inhomogeneous charge distribution set by the thermal equilibrium. This is the same quantity evaluated by Lattice QCD methods in Refs.~\cite{Endrodi:2023wwf}. 

\subsection{Renormalization of electromagnetic susceptibilities}
Both susceptibilities contain an additive divergence, due to a multiplicative ultraviolet (UV) divergence of the electric charge $e$~\cite{Gies:1998vt, DUNNE_2005, Manuel:2016wqs, Carignano:2017, Gorda:2022}. This divergence can be absorbed into the grand canonical potential $\Omega_{\text{\tiny total}}$'s bare vacuum energy 
\begin{equation}
    \Omega_{\text{\tiny total}} \, \supset\, \frac{1}{2}\left(B_b^2-E_b^2\right) \,=\, \frac{1}{2}Z_e \left(B^2-E^2\right) 
\end{equation}
of the electromagnetic field through its wavefunction renormalization, which takes the form~\cite{Baikov:2012zm}
\begin{equation}
    Z_e \,=\, 1-  e^2\left(\, \frac{N_c\vec{Q}^2}{12\pi^2 \epsilon} + \frac{C_F N_c\vec{Q}^2}{32\pi^3 \epsilon}\as  +\mathcal{O}(\as^2)\,\right), 
\end{equation}
where the quark charges enter the expression as
\begin{equation}
\vec{Q}^2\,\equiv\,\sum_f\, \left(\frac{q_f}{e}\right)^2.
\end{equation}
This still leaves freedom in the choice of a renormalization scale $\LambdaIR$, which enters the expressions via the renormalization of the electric charge $e$, while the strong coupling $g$ is accompanied by $\LbarQCD$, the $\overline{\mathrm{MS}}$-scheme scale\footnote{We reserve the symbol $\Lambda$ as a renormalization scale, which is not to be confused with the Landau pole.}:
\begin{equation}
    e_b^2 \,=\, \left(\frac{\Euler^{\gamE}}{4\pi}\right)^{\epsilon}\LambdaIR^{2\epsilon}e^2 Z_e^{-1},\quad g_b^2 \,=\, \left(\frac{\Euler^{\gamE}}{4\pi}\right)^{\epsilon} \LbarQCD^{2\epsilon} 4\pi\as +\mathcal{O}(\as^2) .
\end{equation}
We discuss the choice of $\LambdaIR$ for a consistent comparison to Lattice simulations in Sec.~\ref{Sec:matching}.

\section{Summary of the calculation}
\label{sec:summary-calc}

In this Section, we outline the main steps involved in the computation of the leading $\mathcal{O}(\as)$ hard contribution to the electromagnetic susceptibilities and the LO soft contribution of $\mathcal{O}(\as^{3/2})$. 
We work in the limit of vanishing quark masses and assume equal quark chemical potentials $\muq$ for the $N_f$ quarks.\footnote{At this order, generalization to $N_f$ quarks of different flavors just adds a trivial flavor index to $\muq$ that must be summed over.}
To simplify the notation, we define 
\begin{equation}
\begin{split}
    \mathcal{N}(z)\, &\equiv\, \Psi(z) + \Psi(\overline{z}),  \\
    \mathcal{M}(z) \,&\equiv\, -i\frac{\ud\Psi}{\ud z}(z) + i\frac{\ud\Psi}{\ud z}(\overline{z}), \\
    \mathcal{Z}(z) \,&\equiv\,\zeta(3,z) + \zeta(3,\overline{z}). 
\end{split}
\end{equation}
with $\Psi$ corresponding to the Digamma function and $\zeta$ to the Hurwitz zeta function. These functions carry a dependence with the chemical potential $\muq$ and the temperature $T$ through $z\equiv\frac{1}{2} -i \frac{\muq}{2\pi T}$. The complex conjugate of $z$ is denoted by $\overline{z}$.

\subsection{The LO hard QCD  correction}
At $\mathcal{O}(\as)$, the three diagrams
\begin{equation}
    \begin{tikzpicture}[baseline=-0.09cm]
    \begin{feynman}
    \vertex at (0,0) (f0);
    \vertex at (0.6,0.6) (f1);
    \vertex at (1.2,0) (f2);
    \vertex at (0.6,-0.6) (f3);
    \vertex at (-0.6,0) (p0);
    \vertex at (1.8,0) (p1);
    \diagram*{
    (f0)  --[fermion, quarter left] (f1)  --[fermion, quarter left] (f2)  --[fermion, quarter left] (f3)  --[fermion, quarter left] (f0);
    (f1) --[gluon] (f3);
    {(p0),(f2)} --[photon, momentum = \(k\)] {(f0),(p1)};
    };
    \end{feynman}
    \end{tikzpicture}
\quad + \quad 
    \begin{tikzpicture}[baseline=-0.09cm]
    \begin{feynman}
    \vertex at (0,0) (f0);
    \vertex at (0.6-0.6/1.414,0.6/1.414) (f1);
    \vertex at (1.2,0) (f2);
    \vertex at (0.6+0.6/1.414,0.6/1.414) (f3);
    \vertex at (-0.6,0) (p0);
    \vertex at (1.8,0) (p1);
    \diagram*{
    (f0)  --[fermion, half left,looseness = 1.6] (f2)   --[fermion, half left,looseness = 1.6] (f0);
    (f3) --[gluon, in = -45, out =-135, looseness = 1.15] (f1);
   {(p0),(f2)} --[photon, momentum = \(k\)] {(f0),(p1)};
    };
    \end{feynman}
    \end{tikzpicture}
\quad + \quad 
    \begin{tikzpicture}[baseline=-0.09cm]
    \begin{feynman}
    \vertex at (0,0) (f0);
    \vertex at (0.6-0.6/1.414,-0.6/1.414) (f1);
    \vertex at (1.2,0) (f2);
    \vertex at (0.6+0.6/1.414,-0.6/1.414) (f3);
    \vertex at (-0.6,0) (p0);
    \vertex at (1.8,0) (p1);
    \diagram*{
    (f0)  --[fermion, half left,looseness = 1.6] (f2)   --[fermion, half left,looseness = 1.6] (f0);
    (f3) --[gluon, in = -45+90, out =-135-90, looseness=1.15] (f1);
   {(p0),(f2)} --[photon, momentum = \(k\)] {(f0),(p1)};
    };
    \end{feynman}
    \end{tikzpicture}
\end{equation}
contribute to the photon self energy, from which the susceptibilities are extracted. The second and third diagrams are equal. We evaluate the diagrams by an integration-by-parts reduction~\cite{Laporta_2000,Tarasov_1997} at finite temperature~\cite{Davydychev:2023jto,Nishimura_2012} and density~\cite{_sterman_2023}. The main steps can be summarized as:
\begin{enumerate}
    \item In the static limit $k_0=0$ and in an arbitrary covariant gauge, we evaluate the Dirac traces and simplify the color structure of the transverse and longitudinal components of the photon polarization with the help of \verb|FeynCalc10|~\cite{Shtabovenko_2025,Shtabovenko_2020,Shtabovenko_2016}; 
    
    \item The resulting expression is expanded in the external spatial momentum $k$. Additionally, we rescale each internal momentum with respect to temperature $T$, leaving a dimensionless integration variable for convenience; 
    \item The resulting terms are mapped onto two-loop integrals of the form
    \begin{equation}
        \SumInt_{\{P\},Q}   \frac{p_0^{\eta_1}\, q_0^{\eta_2}\, (P+\hat{k})^{2{\eta_3}} (Q+\hat{k})^{2{\eta_4}} }{P^{2{\eta_5}}\,Q^{2{\eta_6}}\,(P+Q)^{2{\eta_7}}}\, , 
    \end{equation}
    with $\eta_j\in\mathds{Z}$ and $\hat{k}$ a spatial unit vector. This vector is a remnant contribution of $k$ and drops out in the final result. 
    The four-vector $P$ contains the fermionic Matsubara mode $p_0$, which includes the chemical potential shift $-i\mu$, while $Q$ is bosonic with its Matsubara mode denoted by $q_0$. 
    \item The two loop sum-integrals are reduced by \verb|FIRE6.5|~\cite{Mertig:1990an} to a product of one loop sum-integrals. These sum-integrals are solved exactly.
\end{enumerate}
The resulting bare expressions can be found in App.~\ref{Appendix: bare results}. We have verified that the gauge-fixing parameter drops out of our final expressions.

\subsection{The LO soft QCD contribution}
When evaluating the $\mathcal{O}(\as^2)$ contributions to the perturbative expansion, an infrared (IR) divergence appears due to the self-energy corrections of the gluon propagator. To cure this behavior, a resummation of self-energy diagrams along the gluonic lines is needed, leading to an $\mathcal{O}(\as^{3/2})$ EQCD contribution. This can be thought of as gluons becoming sensitive to the soft scale ($\sim g T$), a phenomenon called \textit{Debye screening}~\cite{Blaizot:2001nr,Vuorinen:2003fs, Hietanen:2008tv}. Since fermionic momenta reside at the hard scale ($ \sim 2\pi T$), their corresponding modes can be integrated out when computing such soft contributions. This also allows one to set the soft gluonic momentum $q$ to zero inside the fermion loop in the appropriate diagrams, which in this case are
\begin{equation}\label{eq:EQCD_Operator_matching}
\lim_{q\to0}\ \left\{
\begin{split}
 \begin{tikzpicture}[baseline=-0.07cm]
    \begin{feynman}
    \vertex at (0,0) (f0);
    \vertex at (0.6,0.6) (f1);
    \vertex at (0.6,1.55) (g1) {\(A_0^a\)};
    \vertex at (1.2,0) (f2);
    \vertex at (0.6,-0.6) (f3);
    \vertex at (0.6,-1.55) (g2) {\(A_0^b\)};
    \vertex at (-0.852,0) (p0) {\(\mu\)};
    \vertex at (2.052,0) (p1) {\(\nu\)};
    \diagram*{
    (f0)  --[fermion, quarter left] (f1)  --[fermion, quarter left] (f2)  --[fermion, quarter left] (f3)  --[fermion, quarter left] (f0);
    (f1) --[gluon, reversed momentum' = \(q\)] (g1);
    (f3) --[gluon, momentum = \(q\)] (g2);
    (f2) --[photon, momentum' = \(K\)] (p1);
    (f0) --[photon, reversed momentum = \(K\)] (p0);
    };
    \end{feynman}
    \end{tikzpicture}
 +&
    \begin{tikzpicture}[baseline=-0.07cm]
    \begin{feynman}
    \vertex at (0,0) (f0);
    \vertex at (0.6-0.6/1.414,0.6/1.414) (f1);
    \vertex at (0.6-0.6/1.414,1.55) (g1) {\(A_0^a\)};
    \vertex at (1.2,0) (f2);
    \vertex at (0.6+0.6/1.414,0.6/1.414) (f3);
    \vertex at (0.6+0.6/1.414,1.55) (g2) {\(A_0^b\)};
    \vertex at (-0.852,0) (p0) {\(\mu\)};
    \vertex at (2.052,0) (p1) {\(\nu\)};
    \diagram*{
    (f0)  --[fermion, half left,looseness = 1.6] (f2)   --[fermion, half left,looseness = 1.6] (f0);
    (f1) --[gluon, reversed momentum = \(q\)] (g1);
    (f3) --[gluon, momentum' = \(q\)] (g2);
   {(p0),(f2)} --[photon, momentum' = \(K\)] {(f0),(p1)};
    };
    \end{feynman}
    \end{tikzpicture}
    \\+&
    \begin{tikzpicture}[baseline=-0.07cm]
    \begin{feynman}
    \vertex at (0,0) (f0);
    \vertex at (0.6-0.6/1.414,-0.6/1.414) (f1);
    \vertex at (0.6-0.6/1.414,-1.55) (g1) {\(A_0^a\)};
    \vertex at (1.2,0) (f2);
    \vertex at (0.6+0.6/1.414,-0.6/1.414) (f3);
    \vertex at (0.6+0.6/1.414,-1.55) (g2) {\(A_0^b\)};
    \vertex at (-0.852,0) (p0) {\(\mu\)};
    \vertex at (2.052,0) (p1) {\(\nu\)};
    \diagram*{
    (f0)  --[fermion, half left,looseness = 1.6] (f2)   --[fermion, half left,looseness = 1.6] (f0);
    (f1) --[gluon, reversed momentum' = \(q\)] (g1);
    (f3) --[gluon, momentum = \(q\)] (g2);
   {(p0),(f2)} --[photon, momentum = \(K\)] {(f0),(p1)};
    };
    \end{feynman}
    \end{tikzpicture}
\end{split}\right\}
\quad
=
\  \frac{1}{2!}
\begin{tikzpicture}[baseline=-0.07cm]
\begin{feynman}[small]
\vertex at (0,1.2) (g1) {\(A_0^a\)};
\vertex at (0,-1.2) (g2) {\(A_0^b\)};
\vertex[blob] (x) at (0,0) {};
\diagram*{
(g1) --[gluon] (x) --[gluon] (g2);
};
\end{feynman}
\end{tikzpicture}
\ \equiv\    V_{\mu\nu}(K)\, \delta^{ab}.
\end{equation}
The Lorentz indices $\mu$ and $\nu$ belong to the external photons which carry the momentum $K=(0,k)$ while the EQCD scalar fields $A^a_0$ and $A_0^b$ carry a color index $a$ or $b$. 
Eq.~\eqref{eq:EQCD_Operator_matching} defines a two-point operator, illustrated by the shaded circle and its prefactor $V_{\mu\nu}(k)$. This operator generates an EQCD expectation value $\langle A_0^a A_0^b \rangle$ with respect to the electrostatic modes and quantifies the LO soft contribution to the photon polarization tensor. For the electric or magnetic susceptibility, we extract the corresponding soft scale's contribution from the photon polarization tensor (see App.~\ref{Appendix: bare results} for details) and find
\begin{equation}
\begin{split} 
    \lim_{k\to0} \frac{1}{2(d-1)}\frac{\partial^2\Pi_{ii}}{\partial k^2}(0,k)
    \ &\rightarrow\ \ \
    \lim_{k\to0} \frac{1}{4}\frac{\partial^2 V_{ii}}{\partial k^2}(0,k)
    \delta^{ab} \langle A_0^a A_0^b \rangle\Big|_{d=3}
    = - \frac{\as e^2 \vec{Q}^2  \delta ^{ab}}{24 \pi ^3 T^2}\mathcal{Z}(z)\langle A_0^a A_0^b \rangle\Big|_{d=3}
      ,\\
      -\lim_{k\to0} \frac{1}{2}\frac{\partial^2\Pi_{00}}{\partial k^2}(0,k)
    \ &\rightarrow \,
     -\lim_{k\to0} \frac{1}{2}\frac{\partial^2V_{00}}{\partial k^2}(0,k) \delta^{ab} \langle A_0^a A_0^b \rangle\Big|_{d=3}
    = \ \, \frac{\as e^2 \vec{Q}^2   \delta ^{ab}}{24 \pi ^3 T^2}\mathcal{Z}(z) \langle A_0^a A_0^b \rangle\Big|_{d=3}
      .
\end{split}
\end{equation}
Adding the LO EQCD expectation value yields the result
\begin{equation} 
    \chi\,\Big|_{\mathcal{O}(\as^{3/2})} = \,  -\, \xi\,\Big|_{\mathcal{O}(\as^{3/2})} =  \ N_c C_F\frac{\as m_E  \vec{Q}^2  }{48 \pi ^4 T}\mathcal{Z}(z) ,
\end{equation}
where the EQCD screening mass 
\begin{equation}
\label{eq:screeningmass}
    m_{ \mathrm{E}}^2\,=\,\frac{4\pi \as}{3}\left\{\, \Big( N_c+\frac{1}{2}N_f \Big)\, T^2+\frac{3}{2\pi^2}N_f  \muq^2 \,\right\},
\end{equation}
naturally enters the result.

\subsection{Vacuum contribution matching}\label{Sec:matching}

Finally, to consistently compare our results to Lattice QCD, we must relate $\LbarQED$ to the scale that is determined in Lattice simulations. We do so by matching our expressions to the Lattice in the asymptotic limit $T\rightarrow \infty$. Within Lattice simulations~\cite{Bali:2012jv,Bali:2020bcn,Bonati:2013lca}, a scale $\muQED=115(8)$ MeV~\cite{Bali:2020bcn} appears when extracting the vacuum contribution from the susceptibilities. In terms of this scale, the LO contribution for the magnetic susceptibility can be written as 
\begin{equation}
 \chi_{\mathrm{LO}}\,=\,
 \frac{N_c\vec{Q}^2}{12\pi^2}
 \log \left(\gamma \frac{T^2}{\muQED^2}\right),   
\end{equation}
with  $\gamma$ being a scheme-dependent constant. Therefore, our matching procedure consists of determining $\gamma$ by computing the LO term within LPT and matching it to the LO term of the pQCD expansion.

To determine the renormalized electric and magnetic susceptibilities\footnote{It holds $\chi_b(T=0,\muB=0) = -\xi_b(T=0,\muB=0)$ due to Lorentz invariance, so matching $\LambdaIR$ through $\chi$ is sufficient.} within LPT, we first define it as
\begin{equation}
\begin{split}
    \chi(T) \,&\equiv\, \chi_b(T) - \chi_b(T=0),\\
    \xi(T) \,&\equiv  \ \xi_b(T) - \,\xi_b(T=0),
\end{split}
\end{equation}
at vanishing baryon chemical potential, similarly to Refs.~\cite{Endrodi:2022, Bali:2012jv, Bali:2020bcn}. In full simulations, this condition corresponds to the requirement that the susceptibilities should vanish in vacuum~\cite{Endrodi:2022, Bali:2020bcn}.
The LPT bare susceptibility can be expressed as
\begin{equation}\label{eq:bare_sus_Lattice}
    \chi^{\text{\tiny Lat}}_b \,=\, \frac{  N_c \vec{Q}^2}{12 \pi ^2} \log \left(\frac{ \pi^3 a^2 T^2}{4 \Euler^{2\gamE + 5/4 - 3\pi^2Z/4}}\right) +\mathcal{O}\left(am_f,\frac{m_f}{T}\right)  ,
\end{equation}
with $Z=-0.007638447...$ a numerical constant and $a$ as the Lattice spacing. This is derived in App.~\ref{Appendix:Lattice_pert_theory}. 
While the LO vacuum susceptibility $\chi^{\text{\tiny Lat}}_b(T=0)$ is given by
\begin{equation}\label{eq:bare_Sus_vacuum_Lattice}
    \chi_b^{\text{\tiny Lat}}(T=0) \,=\, \frac{N_c\vec{Q}^2}{12\pi^2} \log\left(a^2 \muQED^2\right),
\end{equation}
with $\muQED=115(8)\,\mathrm{MeV}$ being the constant determined in Ref.~\cite{Bali:2020bcn}.
Combining the bare susceptibility of Eq.~\eqref{eq:bare_sus_Lattice} with the zero temperature susceptibility of Eq.~\eqref{eq:bare_Sus_vacuum_Lattice} yields
\begin{equation}\label{eq:Lattice_Sus}
    \chi^{\text{\tiny Lat}} \,=\, \chi^{\text{\tiny Lat}}_b(T) - \chi_b^{\text{\tiny Lat}}(T=0)\,=\, \frac{  N_c  \vec{Q}^2}{12 \pi ^2} \log \left(\frac{ \pi^3  T^2}{4 \Euler^{2\gamE + 5/4 - 3\pi^2Z/4} \muQED^2}\right),  
\end{equation}
as the regulated susceptibility. At vanishing baryon chemical potential $\muq=0$, we compare the LO magnetic susceptibility in the $\overline{\mathrm{MS}}$-scheme [i.e.~the LO term of Eq.~\eqref{eq:Susceptibilities}] with the LPT result of Eq.~\eqref{eq:Lattice_Sus}. Setting the expressions equal, determining the Lattice scheme $\gamma$ and absorbing $\gamma$ into $\LambdaIR^2 = \pi^2 \Euler^{-2\gamE} \muQED^2 /\gamma$ , we find $\LambdaIR = 249(18)\, \mathrm{MeV}$ as the appropriate $\overline{\mathrm{MS}}$-scale.

\section{Summary of the results}
\label{sec:summary-results}

At finite temperature $T$ and quark chemical potential $\muq$, we summarize our main result as\footnote{We note that the QCD scale $\LbarQCD$ entering at $\mathcal{O}(\as)$ is unusual compared to the QCD pressure. This scale produces an anomalous dimension for the susceptibilities, similar to the anomalous dimensions of the photon polarization tensor~\cite{Baikov:2012zm}. However, the susceptibilities enter the QCD pressure multiplied by the electromagnetic field strength and the electromagnetic vacuum energy. It is this total term which is renormalization-group invariant with a vanishing anomalous dimension.}
\begin{equation}\label{eq:Susceptibilities}
\begin{split} 
    \chi \,=\,&\ \ \, \frac{N_c\vec{Q}^2}{12\pi^2}\Bigg\{  \log\left(\frac{\pi^2T^2}{\LambdaIR^2}\right)+\mathcal{N}(z)+4\log2  +\frac{\as(\Lbar)}{\pi}\frac{C_F}{16}\bigg( 12\,\log\left(\frac{\pi^2T^2}{\LambdaIR \LbarQCD}\right) +48\log2 \\ &+16\gamE  -23 
    +20\, \mathcal{N}(z) + \frac{8\muq}{\pi T}\, \mathcal{M}(z)
    +\Big[1+\frac{\muq^2}{\pi^2 T^2}\Big]\,\mathcal{Z}(z)
    \bigg) +\frac{\as(\Lbar)}{\pi}  C_F \frac{m_\text{E} }{4 \pi  T }  \mathcal{Z}(z) + \mathcal{O}(\as^2)
    \Bigg\} ,\\
     \xi \,=\,& -\frac{N_c\vec{Q}^2}{12\pi^2} \Bigg\{ \log\left(\frac{\pi^2T^2}{\LambdaIR^2}\right)+\mathcal{N}(z)+4\log2 +1
    +\frac{\as(\Lbar)}{\pi}\frac{C_F}{16}\bigg(
    12\,\log\left(\frac{\pi^2T^2}{\LambdaIR \LbarQCD}\right) +48\log2 \\ &-16\gamE  -43 
    +\ 4\, \mathcal{N}(z) + \frac{8\muq}{\pi T}\, \mathcal{M}(z) 
    +\Big[1+\frac{\muq^2}{\pi^2 T^2}\Big]\,\mathcal{Z}(z)
    \bigg) +\frac{\as(\Lbar)}{\pi}  C_F \frac{m_\text{E} }{4 \pi   T }  \mathcal{Z}(z) + \mathcal{O}(\as^2)
    \Bigg\} .
\end{split}
\end{equation}
An interesting first observation is that the logarithmic dependence with the temperature behaves as 
\begin{equation}
\chi\, \Big|_{\log T}\ = \ -\xi\, \Big|_{\log T}\ = \ \frac{N_c \vec{Q}^2 }{12 \pi^2}\left( 1 + \frac{\as(\Lbar)}{\pi}\frac{3C_F}{4}\right)\log\left(T^2\right). 
\end{equation}
The coefficient in parentheses is simply the QED beta function $\beta_{\text{\tiny QED}} $ with QCD corrections~\cite{Baikov:2012zm}. We therefore explicitly verify the conjecture made in Refs.~\cite{Bali:2014kia,Bali:2020bcn,Endrodi:2023wwf} for the high temperature behavior of the electromagnetic susceptibilities.

\begin{figure}[t]
\centering
\includegraphics[width=0.49\textwidth]{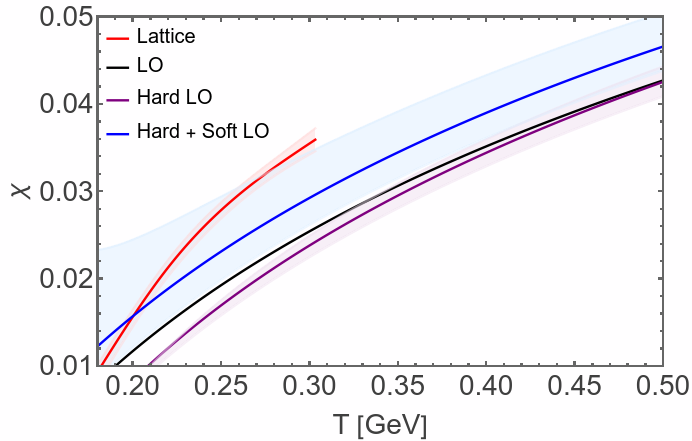}
\includegraphics[width=0.49\textwidth]{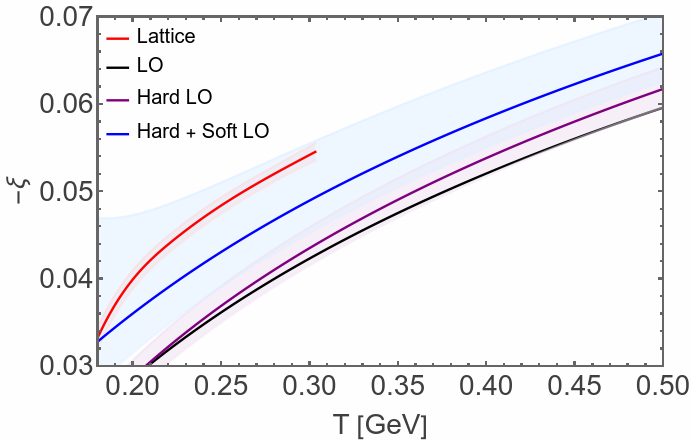}
\caption{The NLO magnetic (left) and electric (right) susceptibility compared to Lattice data with the subtraction scale $\LambdaIR = 249(18)\, \mathrm{MeV}$ set by Lattice perturbation theory. }
\label{fig:Sus_Lattice}
\end{figure}

In Fig.~\ref{fig:Sus_Lattice}, we compare our results in Eqs.~\eqref{eq:Susceptibilities}, taking $\muB=0$, with the Lattice data of Refs.~\cite{Bali:2020bcn,Endrodi:2023wwf}. We consider the three flavor case $N_f=3$, taking a reference value of ${\as(2\,\mathrm{GeV}) = 0.2994}$~\cite{ParticleDataGroup:2008zun}. Additionally, we use three-loop running of the strong coupling $\as$~\cite{Larin_1993,Tarasov:1980au} with a thermal scale of $\Lbar \equiv X \sqrt{(2 \muq)^2 + (0.723 \times 4 \pi T)^2}$ \cite{Huang_1995,Kajantie:1997tt,Kajantie:2002wa,Kurkela:2009gj,Kurkela:2016was,Gorda:2021gha} and $X \in [1/2, 2]$ as the scale variation. We take $\LambdaIR = 249(18)\, \mathrm{MeV}$, as discussed in Sec.~\ref{Sec:matching} and App.~\ref{Appendix:Lattice_pert_theory}. Our plots also account for the statistical error in $\LambdaIR$, which produces a sizable effect.

The full result, which includes the LO QCD correction and the LO EQCD contribution, agrees reasonably well with the Lattice data. We note that the Lattice data are only available up to $T \lesssim 300\,\text{MeV}$ and we expect the agreement to improve at higher temperatures.
The LO hard contribution alone (purple curve) would be insufficient to gain an agreement with the Lattice, while the addition of the EQCD contribution (blue curve) accounts for most of the result's magnitude and is therefore essential.

\begin{figure}[t]
    \centering
    \includegraphics[width=0.49\textwidth]{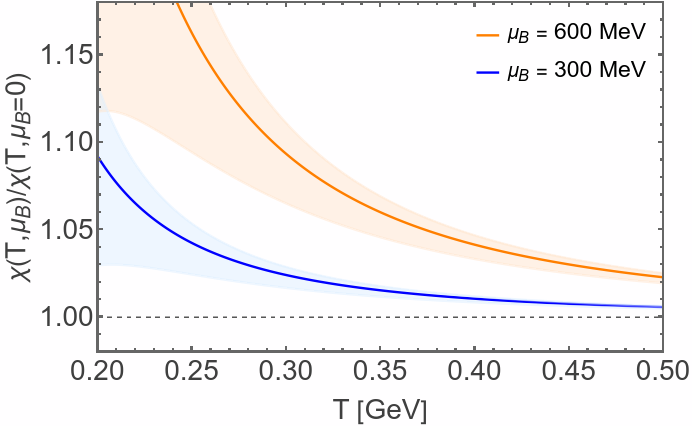}
    \includegraphics[width=0.49\textwidth]{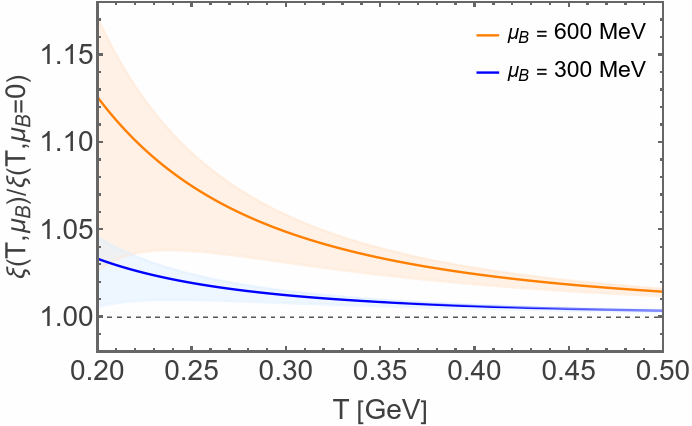}
    \caption{The magnetic (left) and electric (right) susceptibility at finite $\muB$ divided by the corresponding susceptibility with vanishing $\muB$ for $\muB=300\,\mathrm{MeV}$ in blue and $\muB=600\,\mathrm{MeV}$ in orange as a function of temperature. }
    \label{fig:Sus_finite_muB}
\end{figure}

In Fig.~\ref{fig:Sus_finite_muB}, we compare the susceptibilities over a wide range of temperatures at finite values of baryon chemical potential $\muB = 3 \muq$ to the susceptibility with vanishing baryon chemical potential $\muB=0$. The $\muB=300\,\mathrm{MeV}$ and $\muB=600\,\mathrm{MeV}$ results suggest that the susceptibilities $\chi$ and $-\xi$ are increasing with baryon chemical potential and that the magnetic susceptibility is slightly more sensitive to $\muB$.
Towards the cold-and-dense case ($T\to0$ and $\muB>0$), we observe in our results a $\log T$ divergence, signaling the breakdown of dimensional reduction. 
Hence, our perturbative results are only applicable in a region of the phase-diagram which satisfies $2\pi T\gtrsim m_{ \mathrm{E}}$~\cite{Kurkela:2016was} in addition to the weak coupling regime.

\section{Conclusion and future perspectives}
\label{conclusions}

We evaluated the first hard and soft QCD corrections to the static linear response of hot QCD matter to background electric and magnetic fields, which are of order $\mathcal{O}(\as)$ and $\mathcal{O}(\as^{3/2})$, respectively. We showed that, {\it when both contributions are taken into account}, a reasonable agreement with Lattice QCD~\cite{Bali:2013owa, Bali:2020bcn, Endrodi:2023wwf} can be achieved. To draw a consistent comparison to Lattice simulations, we performed a matching between the pQCD expansion and LPT at the asymptotic limit.

Additionally, we explicitly verified up to $\mathcal{O}(\as)$ the claim of Ref.~\cite{Bali:2020bcn} that the coefficient of the leading $\log T$ is proportional to the QED beta function with respect to the strong coupling. 
However, we notice that the soft contributions introduce relevant temperature dependencies that were not anticipated in these early estimates.

It should also be stressed that, despite the observed agreement with Lattice simulations, the available data from Refs.~\cite{Bali:2013owa, Bali:2020bcn, Endrodi:2023wwf} can only reach temperatures up to $\lesssim 300$ MeV, where pQCD may not be reliable. This is reflected in the large scale variation one observes in Figs.~\ref{fig:Sus_Lattice} and \ref{fig:Sus_finite_muB}. Therefore, making a more robust comparison to Lattice simulations requires future investigation, pushing the current perturbative analysis to higher orders and Lattice data to higher temperatures. In this regard, we note that, if the relation between the leading log temperature dependence and the QED beta function holds at higher orders, the $\mathcal{O}(\as^2)$ term is expected to give a positive contribution to $\chi$ and $-\xi$. We believe that this will improve the agreement with Lattice data. 

Finally, our results also indicate that the magnetic and electric responses of QCD matter may be amplified by finite chemical potentials. This motivates further investigations on the electromagnetic response of hot and dense mediums for which pQCD is a crucial source of first-principle information.

The evaluation of the susceptibilities up to and including $\mathcal{O}(\as^{5/2})$ would require three-loop QCD diagrams with mass dimension zero at finite temperature and density. In addition, further operator expectation values within EQCD would be needed, which would account for an external soft momentum expansion.
Notice also that the non-perturbative contribution to the susceptibilities from Magnetostatic QCD (MQCD) sets in at a higher order compared to the equation of state and is of $\mathcal{O}(\as^{7/2})$, see App.~\ref{app-MQCD}. One should also include the strange quark mass effects, possibly a higher order matching to the vacuum contribution and a HTL-resummation to cure the $\log(T)$ IR-divergence. We expect that the latter would produce an enhancing $\as\log(\as)$ term, relevant to the cold and dense case. We hope to report on these further developments in a future publication.

\acknowledgments
 We are thankful to Gunnar S. Bali, Gergely Endr\H{o}di and Pablo Navarrete for useful comments and suggestions. 
 OF and ESF were partially supported by INCT-FNA (Process No. 464898/2014-5), CAPES (Finance Code 001), CNPq, and FAPERJ. OF was also partially supported by the São Paulo Research Foundation (FAPESP) at the late stages of this project. RP has been supported by the Research Council of Finland grants 347499,
353772, and by the ERC Consolidator Grant ExPertQCD (grant No.~101231521). KS also gratefully acknowledges support from the Albert Einstein Center for Fundamental Physics (AEC) at the University of Bern. 


\appendix

\section{Bare results for the soft and hard LO QCD corrections}\label{Appendix: bare results}
The $d=3-2\epsilon$ dimensional expressions contributing at $\mathcal{O}(\as)$ to the susceptibilities are summarized as
\begin{equation}\label{eq:bare_Sus}
\begin{split}
   \chi_b\Big|_{\mathcal{O}(\as)} =\,& 
    -\frac{(d-4) (d-2) \pi ^{d-7}  \vec{Q}^2 \as N_c C_F T^{2 d-6}\Lambda_{\text{\tiny QED}}^{2\epsilon} \Lambda_{\text{\tiny QCD}}^{2\epsilon} \csc ^2\left(\frac{\pi  d}{2}\right)}{384 (d-7) (d-5) d \, \Gamma \left(\frac{d}{2}\right)^2} 
    \Big[\\ &\quad 2 (d-6) (d-5) (d-1)^2 M(6-d,z)(M(2-d,z)\\ &-2 \zeta (2-d))-96 (d-3) \zeta (4-d) M(4-d,z)\\ &+3 (d-3) (d^3-12d^2+41d-14) M(4-d,z)^2\\ &-12
   (d^3-10d+29d-16) M(3-d,z) M(5-d,z)\Big] , \\
   \xi_b\Big|_{\mathcal{O}(\as)}=\,& \ \ 
     \frac{(d-2) (d-1) \pi ^{d-7} \vec{Q}^2 \as N_c C_F T^{2 d-6}\Lambda_{\text{\tiny QED}}^{2\epsilon} \Lambda_{\text{\tiny QCD}}^{2\epsilon} \csc ^2\left(\frac{\pi  d}{2}\right)}{384 (d-7) (d-5) d \, \Gamma \left(\frac{d}{2}\right)^2} \Big[\\ & \quad 48 (d-4) (d-3) \zeta (4-d) M(4-d,z)\\ &+(d-4) ((d-5) (d^3-10d^2 +15d+6) M(6-d,z) (M(2-d,z)-2 \zeta (2-d))\\ &+2 \left((d-5)^2 (d-1)
   d-48\right) M(3-d,z) M(5-d,z))\\ &-3 (d-3) (d^3-7d^2+2d)+28) M(4-d,z)^2\Big] ,
\end{split}
\end{equation}
where $z= \frac{1}{2} -i \frac{\muq}{2\pi T}$ and the definition
\begin{equation}
    M(s-d,z) \,=\, \zeta(s-d,z)+(-1)^s\zeta(s-d,\overline{z})
\end{equation}
holds. Further, the $\overline{\mathrm{MS}}$-scale $\LambdaIR$ enters by its $\mathrm{MS}$ analog $\Lambda_{\text{\tiny QED}}^2 = \frac{\Euler^{\gamE}}{4\pi} \LambdaIR^2 $.

For the EQCD matching coefficients $V_{\mu\nu}(K)$ in $d$ spatial dimensions, it holds
\begin{equation}
\begin{split}
     \lim_{k\to0} \frac{1}{2(d-1)}\frac{\partial^2 V_{ii}}{\partial k^2}(0,k) \,&= 
     \ \ e^2 \as T^{d-5} \Lambda_{\text{\tiny QED}}^{2\epsilon}
     \frac{(d-5) \pi ^{\frac{d}{2}-4}  \csc \left(\frac{\pi  d}{2}\right)}{24 \Gamma \left(\frac{d}{2}-2\right)}M(6-d,z), \\
     -\lim_{k\to0} \frac{1}{2}\frac{\partial^2V_{00}}{\partial k^2} (0,k) \,&=
     -e^2 \as T^{d-5} \Lambda_{\text{\tiny QED}}^{2\epsilon}
     \frac{(d-5) (d-1) \pi ^{\frac{d}{2}-4}  \csc \left(\frac{\pi  d}{2}\right) }{48 \Gamma \left(\frac{d}{2}-2\right)}M(6-d,z).
\end{split}
\end{equation}

\section{Lattice perturbation theory }\label{Appendix:Lattice_pert_theory}
In this appendix, we calculate the magnetic susceptibility to LO in Lattice perturbation theory (LPT).
First, we split the bare magnetic susceptibility into a scheme independent thermal contribution and its UV-divergent vacuum contribution
\begin{equation}\label{eq:Lattice_pert_decomposition}
    \chi^{\text{\tiny Lat}}_b \,=\, \chi_{T\neq0} + \chi^{\text{\tiny Lat}}_{T=0} .
\end{equation}
The thermal contribution is expressed as
\begin{equation}
     \chi_{T\neq0} \,=\,\sum_f 
    \frac{ q_f^2 N_c }{ 12 \pi ^2 e^2}\log \left(\frac{\pi ^2 T^2}{\Euler^{2\gamE}m_f^2}\right) +\mathcal{O}\left(\frac{m_f^2}{T^2}\right),
\end{equation}
with a contribution due to all quark flavors and their mass values $m_f$~\cite{Bali:2020bcn,Endrodi:2022}.

Next, we calculate the vacuum contribution in a naive Lattice discretization~\cite{gattringer2009quantum,Capitani_2003}. After simplifying the Dirac structure and taking the external momentum derivative, we find
\begin{equation}
\begin{split}
    \chi^{\text{\tiny Lat}}_{T=0} \,&=\, \lim_{k_1\to0} \frac{1}{2}\frac{\partial^2\Pi_{22}^{T=0}}{\partial k_1^2}(k_0=0,k_1,k_2=0,k_3=0) \\ 
    \,&=\, 2^{-4} \sum_f  q_f^2 N_c  \int_{-\pi}^{\pi}\frac{\ud^4p}{(2\pi)^4}\Bigg[
    -\frac{2 \cos ^2(p_1)}{\left(a^2 m_f^2+\hat{p}^2\right)^2}+\frac{16 \sin ^2(p_1) \cos ^2(p_1) \sin ^2(p_2)}{\left(a^2 m_f^2+\hat{p}^2\right)^4} \Bigg],
\end{split}
\end{equation}
with $\hat{p}^2 = \sum_{i=1}^4 \sin(p_i)^2$ and an inverse factor of $2^4$ entering to account for fermion doubling. A Schwinger parametrization later~\cite{Becher_2002}, the term can be represented as
\begin{equation}\label{eq:Lattice_pert_bare_full}
\begin{split}
    \chi^{\text{\tiny Lat}}_{T=0}
   \,=\, 2^{-4} \sum_f q_f^2 N_c \int_0^\infty \ud t\ F(t,am_f) ,
\end{split}
\end{equation}
with the integrand $F(t,am_f)$ given by
\begin{equation}
\begin{split}
    F(t,am) \,=&\,\frac{-t}{12} \Euler^{-t \left(a^2 m^2+2\right)}   \Big[\,
    3 t^2 I_1^3 I_0 
    +  \left( 2a^4 m^4 t^2+6 a^2 m^2 t^2+6t^2-t\right) I_0^4 \\
    + &\left( 2 a^4 m^4 t^2+4 a^2 m^2 t -3 t^2-4t+4\right) I_1 I_0^3 
    - \left(6 a^2 m^2 t^2 +6 t^2+t\right) I_1^2 I_0^2
    \,\Big] 
\end{split}
\end{equation}
in terms of the modified Bessel functions $I_0 = I_0(t/2)$ and $I_1 = I_1(t/2)$.
Expanding the function $F(t,am)$ up to and including $\Euler^{-t a^2 m^2}\mathcal{O}(1/t)$, yields the expression
\begin{equation}
\begin{split}
    f(t,am) \,=\, 
      -\Euler^{-a^2 m^2 t} \,\Big[\, \frac{a^4 m^4 t}{3 \pi ^2}+\frac{a^4 m^4}{4 \pi ^2 t}+\frac{a^4 m^4}{6 \pi ^2} +\frac{a^2 m^2}{\pi ^2 t}+\frac{4 a^2 m^2}{3 \pi ^2}+\frac{4}{3 \pi ^2 t}\,\Big] .
\end{split}
\end{equation}
We use this expression to evaluate the integral towards the continuum limit by considering
\begin{equation}\label{eq:Lattice_pert_pieces}
\begin{split}
    &\int_0^\infty \ud t \,\Big[F(t,am=0) -  f(t,am=0)\tanh(t)\Big] \,=\, Z , \\
    &\int_0^\infty \ud t\  f(t,am) \tanh(t) \,=\, 
    \frac{4}{3 \pi ^2} \log \left(\frac{a^2 m^2 \pi}{4 \Euler^{5/4}}\right)+O\left(am\right) ,
\end{split}
\end{equation}
where we evaluate $Z= -0.007638447...$ numerically. Combining Eqs.~(\ref{eq:Lattice_pert_bare_full}) and~(\ref{eq:Lattice_pert_pieces}) yields the susceptibilities vacuum contribution. Inserting the vacuum contribution into Eq.~(\ref{eq:Lattice_pert_decomposition}) finally results in Eq.~(\ref{eq:bare_sus_Lattice}), the bare magnetic susceptibility evaluated by LPT.

\section{Non-perturbative MQCD contribution}
\label{app-MQCD}
QCD's Equation of State (EoS) contains a non-perturbative contribution of $\mathcal{O}(\as^3)$~\cite{Linde:1980ts}. Since it is of theoretical interest at which order in $\as$ observables show a non-perturbative contribution, we will elaborate on it here. Overall, we find a $\mathcal{O}(\as^{7/2})$ contribution to the susceptibilities.

To simplify the problem, we will only outline the matching from QCD to EQCD and finally to MQCD in the presence of electromagnetic background fields.
Consider a QCD Lagrangian with additional photon degrees of freedom $B_\mu$ coupled to quarks 
\begin{equation}
    \mathcal{L}_{\text{\tiny QCD+QED}}(x) \,\supset\, \sum_f q_fB_\mu(x) \overline{\psi}_f(x)\gamma^\mu \psi_f(x).
\end{equation}
This interaction generates gluon-photon vertices containing the field strength of the photon fields $B_i$ and $B_0$
\begin{equation}\label{eq:EQCD_matching}
\begin{tikzpicture}[baseline=0.0cm]
\begin{feynman}
\vertex at (-0.4,0.4) (p1);
\vertex at (-0.4,-0.4) (p2);
\vertex at (0.4,0.4) (g1);
\vertex at (0.4,-0.4) (g2);
\vertex at (-1.15,1.15) (p1o) {\(B_\mu\)};
\vertex at (-1.15,-1.15) (p2o) {\(B_\nu\)};
\vertex at (1.15,1.15) (g1o) {\(A_0\)};
\vertex at (1.15,-1.15) (g2o) {\(A_0\)};
\diagram*{
(p1) --[fermion, quarter right] (p2) --[fermion, quarter right] (g2) --[fermion, quarter right] (g1) --[fermion, quarter right] (p1);
(p1) --[photon] (p1o);
(p2) --[photon] (p2o);
(g1) --[gluon] (g1o);
(g2) --[gluon] (g2o);
};
\end{feynman}
\end{tikzpicture}
\ +\ 
\begin{tikzpicture}[baseline=0.0cm]
\begin{feynman}
\vertex at (-0.4,0.4) (g1);
\vertex at (-0.4,-0.4) (p2);
\vertex at (0.4,0.4) (p1);
\vertex at (0.4,-0.4) (g2);
\vertex at (-1.15,1.15) (g1o) {\(A_0\)};
\vertex at (-1.15,-1.15) (p2o) {\(B_\nu\)};
\vertex at (1.15,1.15) (p1o) {\(B_\mu\)};
\vertex at (1.15,-1.15) (g2o) {\(A_0\)};
\diagram*{
(g1) --[fermion, quarter right] (p2) --[fermion, quarter right] (g2) --[fermion, quarter right] (p1) --[fermion, quarter right] (g1);
(p1) --[photon] (p1o);
(p2) --[photon] (p2o);
(g1) --[gluon] (g1o);
(g2) --[gluon] (g2o);
};
\end{feynman}
\end{tikzpicture} \,\sim\,
e^2\as\left(\partial_iB_j - \partial_jB_i  \right)^2A_0^2 \ + \ e^2\as(\partial_i B_0)^2 A_0^2.
\end{equation}
These interaction contribute to EQCD.

Treating the field strengths $\partial_iB_j - \partial_jB_i$ and $\partial_iB_0$ as homogeneous magnetic $B$ and electric $E$ background fields\footnote{An homogeneous electric background introduces an IR-divergence due to the unbounded nature of the $e^2\as B_0^2A_0^2$ operator. This issue can be regulated by considering an oscillating background field~\cite{Endrodi:2022}, extracting the electric susceptibility from the EoS and recovering the homogeneous electric field in the large wavelength limit.} introduces a correction to the EQCD screening mass $m_{\mathrm{E}}^2\supset e^2 \alpha_s B^2 T^{-2} + e^2\alpha_s E^2T^{-2} $, which enters the NLO MQCD coupling~\cite{Giovannangeli:2003ti,Farakos_1994} and thus the non-perturbative 
\begin{equation}\label{eq:MQCD_coupling}
    g_\mathrm{M}^6 \,=\, g_\mathrm{E}^6\left[1- \frac{1}{16}\frac{g_\mathrm{E}^2 N_c}{\pi m_\mathrm{E}}  + \mathcal{O}\left(\left(\frac{g_\mathrm{E}^2 N_c}{\pi m_\mathrm{E}}\right)^2\right)\right]
\end{equation}
order to the EoS.
Finally, the electric and magnetic susceptibilities can be extracted from the EoS through of Eq.~(\ref{eq:MQCD_coupling}) and $m_{\mathrm{E}}$'s homogeneous background field dependence. 
This leads to an $\mathcal{O}(\as^{7/2})$ MQCD contribution to the susceptibilities.
We also note that the matching we have outlined allows for the precise evaluation of the non-perturbative order by making use of the result of Ref.~\cite{DiRenzo:2006nh}.

\bibliographystyle{JHEP.bst}
\bibliography{references}

@article{Duncan:1992hi,
    author = "Duncan, Robert C. and Thompson, Christopher",
    title = "{Formation of very strongly magnetized neutron stars - implications for gamma-ray bursts}",
    doi = "10.1086/186413",
    journal = "Astrophys. J. Lett.",
    volume = "392",
    pages = "L9",
    year = "1992"
}

@article{Thompson:1993hn,
    author = "Thompson, Christopher and Duncan, Robert C.",
    title = "{Neutron star dynamos and the origins of pulsar magnetism}",
    doi = "10.1086/172580",
    journal = "Astrophys. J.",
    volume = "408",
    pages = "194",
    year = "1993"
}

@article{Kouveliotou:1998ze,
    author = "Kouveliotou, C. and others",
    title = "{An X-ray pulsar with a superstrong magnetic field in the soft gamma-ray repeater SGR 1806-20.}",
    doi = "10.1038/30410",
    journal = "Nature",
    volume = "393",
    pages = "235--237",
    year = "1998"
}

@article{Vachaspati:1991nm,
    author = "Vachaspati, T.",
    title = "{Magnetic fields from cosmological phase transitions}",
    doi = "10.1016/0370-2693(91)90051-Q",
    journal = "Phys. Lett. B",
    volume = "265",
    pages = "258--261",
    year = "1991"
}

@article{Enqvist:1993np,
    author = "Enqvist, K. and Olesen, P.",
    title = "{On primordial magnetic fields of electroweak origin}",
    eprint = "hep-ph/9308270",
    archivePrefix = "arXiv",
    reportNumber = "NBI-HE-93-33",
    doi = "10.1016/0370-2693(93)90799-N",
    journal = "Phys. Lett. B",
    volume = "319",
    pages = "178--185",
    year = "1993"
}

@article{Aguilera-Miret:2025nts,
    author = "Aguilera-Miret, Ricard and Christian, Jan-Erik and Rosswog, Stephan and Palenzuela, Carlos",
    title = "{Robustness of Magnetic Field Amplification in Neutron Star Mergers}",
    eprint = "2504.10604",
    archivePrefix = "arXiv",
    primaryClass = "astro-ph.HE",
    doi = "10.1093/mnras/staf1291",
    journal = "Mon. Not. Roy. Astron. Soc.",
    volume = "3067",
    pages = "3077",
    year = "2025"
}

@article{Kharzeev:2007jp,
    author = "Kharzeev, Dmitri E. and McLerran, Larry D. and Warringa, Harmen J.",
    title = "{The Effects of topological charge change in heavy ion collisions: 'Event by event P and CP violation'}",
    eprint = "0711.0950",
    archivePrefix = "arXiv",
    primaryClass = "hep-ph",
    doi = "10.1016/j.nuclphysa.2008.02.298",
    journal = "Nucl. Phys. A",
    volume = "803",
    pages = "227--253",
    year = "2008"
}

@article{Bzdak:2011yy,
    author = "Bzdak, Adam and Skokov, Vladimir",
    title = "{Event-by-event fluctuations of magnetic and electric fields in heavy ion collisions}",
    eprint = "1111.1949",
    archivePrefix = "arXiv",
    primaryClass = "hep-ph",
    reportNumber = "BNL-96541-2011-JA, RBRC-927",
    doi = "10.1016/j.physletb.2012.02.065",
    journal = "Phys. Lett. B",
    volume = "710",
    pages = "171--174",
    year = "2012"
}

@article{Deng:2012pc,
    author = "Deng, Wei-Tian and Huang, Xu-Guang",
    title = "{Event-by-event generation of electromagnetic fields in heavy-ion collisions}",
    eprint = "1201.5108",
    archivePrefix = "arXiv",
    primaryClass = "nucl-th",
    doi = "10.1103/PhysRevC.85.044907",
    journal = "Phys. Rev. C",
    volume = "85",
    pages = "044907",
    year = "2012"
}

@article{Inghirami:2016iru,
    author = "Inghirami, Gabriele and Del Zanna, Luca and Beraudo, Andrea and Moghaddam, Mohsen Haddadi and Becattini, Francesco and Bleicher, Marcus",
    title = "{Numerical magneto-hydrodynamics for relativistic nuclear collisions}",
    eprint = "1609.03042",
    archivePrefix = "arXiv",
    primaryClass = "hep-ph",
    doi = "10.1140/epjc/s10052-016-4516-8",
    journal = "Eur. Phys. J. C",
    volume = "76",
    number = "12",
    pages = "659",
    year = "2016"
}

@article{Roy:2017yvg,
    author = "Roy, Victor and Pu, Shi and Rezzolla, Luciano and Rischke, Dirk H.",
    title = "{Effect of intense magnetic fields on reduced-MHD evolution in $\sqrt{s_{\rm NN}}$ = 200 GeV Au+Au collisions}",
    eprint = "1706.05326",
    archivePrefix = "arXiv",
    primaryClass = "nucl-th",
    doi = "10.1103/PhysRevC.96.054909",
    journal = "Phys. Rev. C",
    volume = "96",
    number = "5",
    pages = "054909",
    year = "2017"
}

@article{Tuchin:2013apa,
    author = "Tuchin, Kirill",
    title = "{Time and space dependence of the electromagnetic field in relativistic heavy-ion collisions}",
    eprint = "1305.5806",
    archivePrefix = "arXiv",
    primaryClass = "hep-ph",
    doi = "10.1103/PhysRevC.88.024911",
    journal = "Phys. Rev. C",
    volume = "88",
    number = "2",
    pages = "024911",
    year = "2013"
}

@article{Tuchin:2015oka,
    author = "Tuchin, Kirill",
    title = "{Initial value problem for magnetic fields in heavy ion collisions}",
    eprint = "1508.06925",
    archivePrefix = "arXiv",
    primaryClass = "hep-ph",
    doi = "10.1103/PhysRevC.93.014905",
    journal = "Phys. Rev. C",
    volume = "93",
    number = "1",
    pages = "014905",
    year = "2016"
}

@article{Shen:2025unr,
    author = "Shen, Diyu and Chen, Jinhui and Huang, Xu-Guang and Ma, Yu-Gang and Tang, Aihong and Wang, Gang",
    title = "{A Review of Intense Electromagnetic Fields in Heavy-Ion Collisions: Theoretical Predictions and Experimental Results}",
    doi = "10.34133/research.0726",
    journal = "Research",
    volume = "8",
    pages = "0726",
    year = "2025"
}

@article{Skokov:2009qp,
    author = "Skokov, V. and Illarionov, A. Yu. and Toneev, V.",
    title = "{Estimate of the magnetic field strength in heavy-ion collisions}",
    eprint = "0907.1396",
    archivePrefix = "arXiv",
    primaryClass = "nucl-th",
    doi = "10.1142/S0217751X09047570",
    journal = "Int. J. Mod. Phys. A",
    volume = "24",
    pages = "5925--5932",
    year = "2009"
}

@article{Hands:2007by,
    author = "Hands, Simon",
    editor = "Kunihiro, Teiji and others",
    title = "{Simulating dense matter}",
    eprint = "hep-lat/0703017",
    archivePrefix = "arXiv",
    doi = "10.1143/PTPS.168.253",
    journal = "Prog. Theor. Phys. Suppl.",
    volume = "168",
    pages = "253--260",
    year = "2007"
}

@article{deforcrand2009pos,
  author        = {de Forcrand, Philippe},
  title         = "{Simulating QCD at finite density}",
  journal       = {Proceedings of Science (PoS)},
  volume        = {LAT2009},
  pages         = {010},
  year          = {2009},
  doi           = {10.22323/1.091.0010},
  eprint        = {0905.4268},
  archivePrefix = {arXiv},
  primaryClass  = {hep-lat}
}

@article{Voronyuk:2011jd,
    author = "Voronyuk, V. and Toneev, V. D. and Cassing, W. and Bratkovskaya, E. L. and Konchakovski, V. P. and Voloshin, S. A.",
    title = "{(Electro-)Magnetic field evolution in relativistic heavy-ion collisions}",
    eprint = "1103.4239",
    archivePrefix = "arXiv",
    primaryClass = "nucl-th",
    doi = "10.1103/PhysRevC.83.054911",
    journal = "Phys. Rev. C",
    volume = "83",
    pages = "054911",
    year = "2011"
}

@article{Galatyuk:2019lcf,
    author = "Galatyuk, Tetyana",
    editor = "Antinori, Federico and Dainese, Andrea and Giubellino, Paolo and Greco, Vincenzo and Lombardo, Maria Paola and Scomparin, Enrico",
    title = "{Future facilities for high $\mu_B$ physics}",
    doi = "10.1016/j.nuclphysa.2018.11.025",
    journal = "Nucl. Phys. A",
    volume = "982",
    pages = "163--169",
    year = "2019"
}

@article{Grasso:2000wj,
    author = "Grasso, Dario and Rubinstein, Hector R.",
    title = "{Magnetic fields in the early universe}",
    eprint = "astro-ph/0009061",
    archivePrefix = "arXiv",
    reportNumber = "DFPD-00-TH-35",
    doi = "10.1016/S0370-1573(00)00110-1",
    journal = "Phys. Rept.",
    volume = "348",
    pages = "163--266",
    year = "2001"
}

@article{Samanta:2025xjm,
    author = "Samanta, Rupam and Broniowski, Wojciech",
    title = "{Magnetic susceptibility of hot hadronic medium and quark degrees of freedom near the QCD cross-over point}",
    eprint = "2511.19255",
    archivePrefix = "arXiv",
    primaryClass = "hep-ph",
    month = "11",
    year = "2025"
}

@article{Endrodi:2023wwf,
    author = "Endrodi, Gergely and Marko, Gergely",
    title = "{QCD phase diagram and equation of state in background electric fields}",
    eprint = "2309.07058",
    archivePrefix = "arXiv",
    primaryClass = "hep-lat",
    doi = "10.1103/PhysRevD.109.034506",
    journal = "Phys. Rev. D",
    volume = "109",
    number = "3",
    pages = "034506",
    year = "2024"
}

@article{Endrodi:2022,
    author = "Endr\H{o}di, Gergely and Mark\'o, Gergely",
    title = "{On electric fields in hot QCD: perturbation theory}",
    eprint = "2208.14306",
    archivePrefix = "arXiv",
    primaryClass = "hep-ph",
    doi = "10.1007/JHEP12(2022)015",
    journal = "JHEP",
    volume = "12",
    pages = "015",
    year = "2022"
}

@article{Ferreira:2023cqw,
    author = "Ferreira, Osvaldo and Fraga, Eduardo S.",
    title = "{Power corrections to the photon polarization tensor in a hot and dense medium of massive fermions}",
    eprint = "2309.06524",
    archivePrefix = "arXiv",
    primaryClass = "hep-ph",
    doi = "10.1103/PhysRevD.109.016025",
    journal = "Phys. Rev. D",
    volume = "109",
    number = "1",
    pages = "016025",
    year = "2024"
}

@article{Endrodi:2026kmb,
    author = "Endr{\H{o}}di, Gergely and Mark{\'o}, Gergely and Sandbote, Leon",
    title = "{On electric fields in hot QCD: infrared regularization dependence}",
    eprint = "2601.01478",
    archivePrefix = "arXiv",
    primaryClass = "hep-ph",
    month = "1",
    year = "2026"
}

@article{Bali:2012jv,
    author = "Bali, G. S. and Bruckmann, F. and Constantinou, M. and Costa, M. and Endrodi, G. and Katz, S. D. and Panagopoulos, H. and Schafer, A.",
    title = "{Magnetic susceptibility of QCD at zero and at finite temperature from the lattice}",
    eprint = "1209.6015",
    archivePrefix = "arXiv",
    primaryClass = "hep-lat",
    doi = "10.1103/PhysRevD.86.094512",
    journal = "Phys. Rev. D",
    volume = "86",
    pages = "094512",
    year = "2012"
}

@article{Bali:2014kia,
    author = {Bali, G. S. and Bruckmann, F. and Endr\"odi, G. and Katz, S. D. and Sch\"afer, A.},
    title = "{The QCD equation of state in background magnetic fields}",
    eprint = "1406.0269",
    archivePrefix = "arXiv",
    primaryClass = "hep-lat",
    doi = "10.1007/JHEP08(2014)177",
    journal = "JHEP",
    volume = "08",
    pages = "177",
    year = "2014"
}

@article{Bali:2020bcn,
    author = "Bali, Gunnar S. and Endr\H{o}di, Gergely and Piemonte, Stefano",
    title = "{Magnetic susceptibility of QCD matter and its decomposition from the lattice}",
    eprint = "2004.08778",
    archivePrefix = "arXiv",
    primaryClass = "hep-lat",
    doi = "10.1007/JHEP07(2020)183",
    journal = "JHEP",
    volume = "07",
    pages = "183",
    year = "2020"
}

@article{Bonati:2013lca,
    author = "Bonati, Claudio and D'Elia, Massimo and Mariti, Marco and Negro, Francesco and Sanfilippo, Francesco",
    title = "{Magnetic Susceptibility of Strongly Interacting Matter across the Deconfinement Transition}",
    eprint = "1307.8063",
    archivePrefix = "arXiv",
    primaryClass = "hep-lat",
    reportNumber = "IFUP-TH-2013-15",
    doi = "10.1103/PhysRevLett.111.182001",
    journal = "Phys. Rev. Lett.",
    volume = "111",
    pages = "182001",
    year = "2013"
}

@article{Steinert:2013fza,
    author = "Steinert, T. and Cassing, W.",
    title = "{Electric and magnetic response of hot QCD matter}",
    eprint = "1312.3189",
    archivePrefix = "arXiv",
    primaryClass = "hep-ph",
    doi = "10.1103/PhysRevC.89.035203",
    journal = "Phys. Rev. C",
    volume = "89",
    number = "3",
    pages = "035203",
    year = "2014"
}

@article{Tawfik:2017cdx,
    author = "Tawfik, Abdel Nasser and Diab, Abdel Magied and Hussein, M. T.",
    title = "{SU(3) Polyakov linear-sigma model: Magnetic properties of QCD matter in thermal and dense medium}",
    eprint = "1712.03264",
    archivePrefix = "arXiv",
    primaryClass = "hep-ph",
    reportNumber = "ECTP-2016-11, WLCAPP-2016-11",
    doi = "10.1134/S1063776118050138",
    journal = "J. Exp. Theor. Phys.",
    volume = "126",
    number = "5",
    pages = "620--632",
    year = "2018"
}

@article{Bali:2013owa,
    author = "Bali, G. S. and Bruckmann, F. and Endrodi, G. and Schafer, A.",
    title = "{Paramagnetic squeezing of QCD matter}",
    eprint = "1311.2559",
    archivePrefix = "arXiv",
    primaryClass = "hep-lat",
    doi = "10.1103/PhysRevLett.112.042301",
    journal = "Phys. Rev. Lett.",
    volume = "112",
    pages = "042301",
    year = "2014"
}

@article{Pang:2016yuh,
    author = "Pang, Long-Gang and Endr\H{o}di, Gergely and Petersen, Hannah",
    title = "{Magnetic-field-induced squeezing effect at energies available at the BNL Relativistic Heavy Ion Collider and at the CERN Large Hadron Collider}",
    eprint = "1602.06176",
    archivePrefix = "arXiv",
    primaryClass = "nucl-th",
    doi = "10.1103/PhysRevC.93.044919",
    journal = "Phys. Rev. C",
    volume = "93",
    number = "4",
    pages = "044919",
    year = "2016"
}

@article{Jiang:2024bez,
    author = "Jiang, Ze-Fang and Zhang, Zi-Han and Yuan, Xue-Fei and Zhang, Ben-Wei",
    title = "{External-magnetic-field-induced paramagnetic squeezing effect in heavy-ion collisions at energies available at the CERN Large Hadron Collider}",
    eprint = "2405.02610",
    archivePrefix = "arXiv",
    primaryClass = "nucl-th",
    doi = "10.1103/PhysRevC.110.014902",
    journal = "Phys. Rev. C",
    volume = "110",
    number = "1",
    pages = "014902",
    year = "2024"
}

@article{Baikov:2012zm,
    author = "Baikov, P. A. and Chetyrkin, K. G. and Kuhn, J. H. and Rittinger, J.",
    title = "{Vector Correlator in Massless QCD at Order $\mathcal{O}(\alpha^4_s)$ and the QED beta-function at Five Loop}",
    eprint = "1206.1284",
    archivePrefix = "arXiv",
    primaryClass = "hep-ph",
    reportNumber = "SFB-CPP-12-36, TTP12-018",
    doi = "10.1007/JHEP07(2012)017",
    journal = "JHEP",
    volume = "07",
    pages = "017",
    year = "2012"
}

@article{Larin_1993,
   title={The three-loop QCD $\beta$-function and anomalous dimensions},
   volume={303},
   ISSN={0370-2693},
   url={http://dx.doi.org/10.1016/0370-2693(93)91441-O},
   DOI={10.1016/0370-2693(93)91441-o},
   number={3-4},
   journal={Physics Letters B},
   publisher={Elsevier BV},
   author={Larin, S.A. and Vermaseren, J.A.M.},
   year={1993},
   month=Apr, pages={334–336} }

@article{Tarasov:1980au,
    author = "Tarasov, O. V. and Vladimirov, A. A. and Zharkov, A. Yu.",
    title = "{The Gell-Mann-Low Function of QCD in the Three Loop Approximation}",
    doi = "10.1016/0370-2693(80)90358-5",
    journal = "Phys. Lett. B",
    volume = "93",
    pages = "429--432",
    year = "1980"
}

@article{Weldon:1982,
    author = "Weldon, H. Arthur",
    title = "{Covariant Calculations at Finite Temperature: The Relativistic Plasma}",
    reportNumber = "PRINT-82-0313 (PENN)",
    doi = "10.1103/PhysRevD.26.1394",
    journal = "Phys. Rev. D",
    volume = "26",
    pages = "1394",
    year = "1982"
}

@article{Blaizot:2001nr,
    author = "Blaizot, Jean-Paul and Iancu, Edmond",
    title = "{The Quark gluon plasma: Collective dynamics and hard thermal loops}",
    eprint = "hep-ph/0101103",
    archivePrefix = "arXiv",
    reportNumber = "SACLAY-T01-005, CERN-TH-2000-272",
    doi = "10.1016/S0370-1573(01)00061-8",
    journal = "Phys. Rept.",
    volume = "359",
    pages = "355--528",
    year = "2002"
}

@article{Gorda:2022,
    author = {Gorda, Tyler and Kurkela, Aleksi and \"Osterman, Juuso and Paatelainen, Risto and S\"appi, Saga and Schicho, Philipp and Sepp\"anen, Kaapo and Vuorinen, Aleksi},
    title = "{Soft photon propagation in a hot and dense medium to next-to-leading order}",
    eprint = "2204.11279",
    archivePrefix = "arXiv",
    primaryClass = "hep-ph",
    reportNumber = "HIP-2022-8/TH",
    doi = "10.1103/PhysRevD.107.036012",
    journal = "Phys. Rev. D",
    volume = "107",
    number = "3",
    pages = "036012",
    year = "2023"
}

@article{Carignano:2017,
    author = "Carignano, Stefano and Manuel, Cristina and Soto, Joan",
    title = "{Power corrections to the HTL effective Lagrangian of QED}",
    eprint = "1712.07949",
    archivePrefix = "arXiv",
    primaryClass = "hep-ph",
    doi = "10.1016/j.physletb.2018.03.012",
    journal = "Phys. Lett. B",
    volume = "780",
    pages = "308--312",
    year = "2018"
}

@article{Manuel:2016wqs,
    author = "Manuel, Cristina and Soto, Joan and Stetina, Stephan",
    title = "{On-shell effective field theory: A systematic tool to compute power corrections to the hard thermal loops}",
    eprint = "1603.05514",
    archivePrefix = "arXiv",
    primaryClass = "hep-ph",
    doi = "10.1103/PhysRevD.94.025017",
    journal = "Phys. Rev. D",
    volume = "94",
    number = "2",
    pages = "025017",
    year = "2016",
    note = "[Erratum: Phys.Rev.D 96, 129901 (2017)]"
}

@article{Gies:1998vt,
    author = "Gies, Holger",
    title = "{QED effective action at finite temperature}",
    eprint = "hep-ph/9812436",
    archivePrefix = "arXiv",
    doi = "10.1103/PhysRevD.60.105002",
    journal = "Phys. Rev. D",
    volume = "60",
    pages = "105002",
    year = "1999"
}

@article{ParticleDataGroup:2008zun,
    author = "Amsler, Claude and others",
    collaboration = "Particle Data Group",
    title = "{Review of Particle Physics}",
    doi = "10.1016/j.physletb.2008.07.018",
    journal = "Phys. Lett. B",
    volume = "667",
    pages = "1--1340",
    year = "2008"
}

@article{Kajantie:1997tt,
    author = "Kajantie, K. and Laine, M. and Rummukainen, K. and Shaposhnikov, Mikhail E.",
    title = "{3-D SU(N) + adjoint Higgs theory and finite temperature QCD}",
    eprint = "hep-ph/9704416",
    archivePrefix = "arXiv",
    reportNumber = "CERN-TH-97-081, CERN-TH-97-81, BI-TP-97-10, HD-THEP-97-17",
    doi = "10.1016/S0550-3213(97)00425-2",
    journal = "Nucl. Phys. B",
    volume = "503",
    pages = "357--384",
    year = "1997"
}

@article{Kajantie:2002wa,
    author = "Kajantie, K. and Laine, M. and Rummukainen, K. and Schroder, Y.",
    title = "{The Pressure of hot QCD up to g6 ln(1/g)}",
    eprint = "hep-ph/0211321",
    archivePrefix = "arXiv",
    reportNumber = "CERN-TH-2002-334, HIP-2002-62-TH, MIT-CTP-3325",
    doi = "10.1103/PhysRevD.67.105008",
    journal = "Phys. Rev. D",
    volume = "67",
    pages = "105008",
    year = "2003"
}

@article{Kurkela:2009gj,
    author = "Kurkela, Aleksi and Romatschke, Paul and Vuorinen, Aleksi",
    title = "{Cold Quark Matter}",
    eprint = "0912.1856",
    archivePrefix = "arXiv",
    primaryClass = "hep-ph",
    reportNumber = "BI-TP-2009-30, CERN-PH-TH-2009-229, INT-PUB-09-060, TUW-09-19",
    doi = "10.1103/PhysRevD.81.105021",
    journal = "Phys. Rev. D",
    volume = "81",
    pages = "105021",
    year = "2010"
}

@article{Kurkela:2016was,
    author = "Kurkela, Aleksi and Vuorinen, Aleksi",
    title = "{Cool quark matter}",
    eprint = "1603.00750",
    archivePrefix = "arXiv",
    primaryClass = "hep-ph",
    reportNumber = "HIP-2016-06-TH, CERN-TH-2016-045",
    doi = "10.1103/PhysRevLett.117.042501",
    journal = "Phys. Rev. Lett.",
    volume = "117",
    number = "4",
    pages = "042501",
    year = "2016"
}

@article{Gorda:2021gha,
    author = {Gorda, Tyler and S\"appi, Saga},
    title = "{Cool quark matter with perturbative quark masses}",
    eprint = "2112.11472",
    archivePrefix = "arXiv",
    primaryClass = "hep-ph",
    doi = "10.1103/PhysRevD.105.114005",
    journal = "Phys. Rev. D",
    volume = "105",
    number = "11",
    pages = "114005",
    year = "2022"
}

@article{Huang_1995,
   title={The relevant scale parameter in the high temperature phase of QCD},
   volume={438},
   ISSN={0550-3213},
   url={http://dx.doi.org/10.1016/0550-3213(95)00007-F},
   DOI={10.1016/0550-3213(95)00007-f},
   number={1–2},
   journal={Nuclear Physics B},
   publisher={Elsevier BV},
   author={Huang, Suzhou and Lissia, Marcello},
   year={1995},
   month=mar, pages={54–66} }

@article{Davydychev:2023jto,
    author = {Davydychev, Andrei I. and Navarrete, Pablo and Schr{\"o}der, York},
    title = "{Factorizing two-loop vacuum sum-integrals}",
    eprint = "2312.17367",
    archivePrefix = "arXiv",
    primaryClass = "hep-ph",
    doi = "10.1007/JHEP02(2024)104",
    journal = "JHEP",
    volume = "02",
    pages = "104",
    year = "2024"
}

@article{Laporta_2000, volume={15},
    title = "{High-precision calculation of multi-loop Feynman integrals by difference equations}",
   ISSN={0217-751X},
   url={http://dx.doi.org/10.1016/S0217-751X(00)00215-7},
   DOI={10.1016/s0217-751x(00)00215-7},
   journal={International Journal of Modern Physics A},
   publisher={World Scientific Pub Co Pte Lt},
   author={Laporta},
   year={2000},
   pages={5087} }

@article{Nishimura_2012,
   title={IBP methods at finite temperature},
   volume={2012},
   ISSN={1029-8479},
   url={http://dx.doi.org/10.1007/JHEP09(2012)051},
   DOI={10.1007/jhep09(2012)051},
   number={9},
   journal={Journal of High Energy Physics},
   publisher={Springer Science and Business Media LLC},
   author={Nishimura, Madoka and Schröder, York},
   year={2012},
   month=sep }

@article{Tarasov_1997,
   title={Generalized recurrence relations for two-loop propagator integrals with arbitrary masses},
   volume={502},
   ISSN={0550-3213},
   url={http://dx.doi.org/10.1016/S0550-3213(97)00376-3},
   DOI={10.1016/s0550-3213(97)00376-3},
   number={1–2},
   journal={Nuclear Physics B},
   publisher={Elsevier BV},
   author={Tarasov, O.V.},
   year={1997},
   month=sep, pages={455–482} }

@article{_sterman_2023,
   title={Integrating by parts at finite density},
   volume={2023},
   ISSN={1029-8479},
   url={http://dx.doi.org/10.1007/JHEP08(2023)212},
   DOI={10.1007/jhep08(2023)212},
   number={8},
   journal={Journal of High Energy Physics},
   publisher={Springer Science and Business Media LLC},
   author={Österman, Juuso and Schicho, Philipp and Vuorinen, Aleksi},
   year={2023},
   month=aug }

@article{Shtabovenko_2025,
   title={FeynCalc 10: Do multiloop integrals dream of computer codes?},
   volume={306},
   ISSN={0010-4655},
   url={http://dx.doi.org/10.1016/j.cpc.2024.109357},
   DOI={10.1016/j.cpc.2024.109357},
   journal={Computer Physics Communications},
   publisher={Elsevier BV},
   author={Shtabovenko, Vladyslav and Mertig, Rolf and Orellana, Frederik},
   year={2025},
   month=jan, pages={109357} }

@article{Shtabovenko_2020,
   title={FeynCalc 9.3: New features and improvements},
   volume={256},
   ISSN={0010-4655},
   url={http://dx.doi.org/10.1016/j.cpc.2020.107478},
   DOI={10.1016/j.cpc.2020.107478},
   journal={Computer Physics Communications},
   publisher={Elsevier BV},
   author={Shtabovenko, Vladyslav and Mertig, Rolf and Orellana, Frederik},
   year={2020},
   month=nov, pages={107478} }

@article{Shtabovenko_2016,
   title={New developments in FeynCalc 9.0},
   volume={207},
   ISSN={0010-4655},
   url={http://dx.doi.org/10.1016/j.cpc.2016.06.008},
   DOI={10.1016/j.cpc.2016.06.008},
   journal={Computer Physics Communications},
   publisher={Elsevier BV},
   author={Shtabovenko, Vladyslav and Mertig, Rolf and Orellana, Frederik},
   year={2016},
   month=oct, pages={432–444} }

@article{Mertig:1990an,
    author = "Mertig, R. and Bohm, M. and Denner, Ansgar",
    title = "{FEYN CALC: Computer algebraic calculation of Feynman amplitudes}",
    reportNumber = "PRINT-90-0639 (WURZBURG)",
    doi = "10.1016/0010-4655(91)90130-D",
    journal = "Comput. Phys. Commun.",
    volume = "64",
    pages = "345--359",
    year = "1991"
}

@article{Becher_2002,
   title={Asymptotic expansion of lattice loop integrals around the continuum limit},
   volume={66},
   ISSN={1089-4918},
   url={http://dx.doi.org/10.1103/PhysRevD.66.074508},
   DOI={10.1103/physrevd.66.074508},
   number={7},
   journal={Physical Review D},
   publisher={American Physical Society (APS)},
   author={Becher, Thomas and Melnikov, Kirill},
   year={2002},
   month=oct }

@book{gattringer2009quantum,
  title={Quantum chromodynamics on the lattice: an introductory presentation},
  author={Gattringer, Christof and Lang, Christian},
  volume={788},
  year={2009},
  publisher={Springer Science \& Business Media}
}

@article{Capitani_2003,
   title={Lattice perturbation theory},
   volume={382},
   ISSN={0370-1573},
   url={http://dx.doi.org/10.1016/S0370-1573(03)00211-4},
   DOI={10.1016/s0370-1573(03)00211-4},
   number={3–5},
   journal={Physics Reports},
   publisher={Elsevier BV},
   author={Capitani, S},
   year={2003},
   month=jul, pages={113–302} }

@inbook{DUNNE_2005,
   title={HEISENBERG–EULER EFFECTIVE LAGRANGIANS: BASICS AND EXTENSIONS},
   ISBN={9789812775344},
   url={http://dx.doi.org/10.1142/9789812775344_0014},
   DOI={10.1142/9789812775344_0014},
   booktitle={From Fields to Strings: Circumnavigating Theoretical Physics},
   publisher={WORLD SCIENTIFIC},
   author={DUNNE, GERALD V.},
   year={2005},
   month=feb, pages={445–522} }

@article{Linde:1980ts,
    author = "Linde, Andrei D.",
    title = "{Infrared Problem in Thermodynamics of the Yang-Mills Gas}",
    reportNumber = "LEBEDEV-80-106",
    doi = "10.1016/0370-2693(80)90769-8",
    journal = "Phys. Lett. B",
    volume = "96",
    pages = "289--292",
    year = "1980"
}

@article{Giovannangeli:2003ti,
    author = "Giovannangeli, P.",
    title = "{Two loop renormalization of the magnetic coupling in hot QCD}",
    eprint = "hep-ph/0312307",
    archivePrefix = "arXiv",
    doi = "10.1016/j.physletb.2004.02.011",
    journal = "Phys. Lett. B",
    volume = "585",
    pages = "144--148",
    year = "2004"
}

@article{Farakos_1994,
   title={3D physics and the electroweak phase transition: Perturbation theory},
   volume={425},
   ISSN={0550-3213},
   url={http://dx.doi.org/10.1016/0550-3213(94)90173-2},
   DOI={10.1016/0550-3213(94)90173-2},
   number={1-2},
   journal={Nuclear Physics B},
   publisher={Elsevier BV},
   author={Farakos, K. and Kajantie, K. and Rummukainen, K. and Shaposhnikov, M.},
   year={1994},
   month=Aug, pages={67–109} }

@article{DiRenzo:2006nh,
    author = "Di Renzo, F. and Laine, M. and Miccio, V. and Schroder, Y. and Torrero, C.",
    title = "{The Leading non-perturbative coefficient in the weak-coupling expansion of hot QCD pressure}",
    eprint = "hep-ph/0605042",
    archivePrefix = "arXiv",
    reportNumber = "BI-TP-2006-13",
    doi = "10.1088/1126-6708/2006/07/026",
    journal = "JHEP",
    volume = "07",
    pages = "026",
    year = "2006"
}

@article{GINSPARG1980388,
author = {P. Ginsparg},
title = {First and second order phase transitions in gauge theories at finite temperature},
journal = {Nuclear Physics B},
volume = {170},
number = {3},
pages = {388-408},
year = {1980},
issn = {0550-3213},
doi = {https://doi.org/10.1016/0550-3213(80)90418-6},
url = {https://www.sciencedirect.com/science/article/pii/0550321380904186},
}

@article{PhysRevD.27.917,
  title = {Dimensional reduction in finite-temperature quantum chromodynamics},
  author = {Nadkarni, Sudhir},
  journal = {Phys. Rev. D},
  volume = {27},
  issue = {4},
  pages = {917--931},
  numpages = {0},
  year = {1983},
  month = {Feb},
  publisher = {American Physical Society},
  doi = {10.1103/PhysRevD.27.917},
  url = {https://link.aps.org/doi/10.1103/PhysRevD.27.917}
}

@article{PhysRevD.23.2305,
  title = {High-temperature Yang-Mills theories and three-dimensional quantum chromodynamics},
  author = {Appelquist, Thomas and Pisarski, Robert D.},
  journal = {Phys. Rev. D},
  volume = {23},
  issue = {10},
  pages = {2305--2317},
  numpages = {0},
  year = {1981},
  month = {May},
  publisher = {American Physical Society},
  doi = {10.1103/PhysRevD.23.2305},
  url = {https://link.aps.org/doi/10.1103/PhysRevD.23.2305}
}

@article{Hietanen:2008tv,
    author = "Hietanen, A. and Kajantie, K. and Laine, M. and Rummukainen, K. and Schroder, Y.",
    title = "{Three-dimensional physics and the pressure of hot QCD}",
    eprint = "0811.4664",
    archivePrefix = "arXiv",
    primaryClass = "hep-lat",
    reportNumber = "BI-TP-2008-26",
    doi = "10.1103/PhysRevD.79.045018",
    journal = "Phys. Rev. D",
    volume = "79",
    pages = "045018",
    year = "2009"
}

@article{Vuorinen:2003fs,
    author = "Vuorinen, A.",
    title = "{The Pressure of QCD at finite temperatures and chemical potentials}",
    eprint = "hep-ph/0305183",
    archivePrefix = "arXiv",
    reportNumber = "HIP-2003-30-TH",
    doi = "10.1103/PhysRevD.68.054017",
    journal = "Phys. Rev. D",
    volume = "68",
    pages = "054017",
    year = "2003"
}

@article{Nagata:2021ugx,
    author = "Nagata, Keitaro",
    title = "{Finite-density lattice QCD and sign problem: Current status and open problems}",
    eprint = "2108.12423",
    archivePrefix = "arXiv",
    primaryClass = "hep-lat",
    doi = "10.1016/j.ppnp.2022.103991",
    journal = "Prog. Part. Nucl. Phys.",
    volume = "127",
    pages = "103991",
    year = "2022"
}
\end{document}